\documentclass[aps,prb,reprint,superscriptaddress,floatfix]{revtex4-1}
\usepackage[utf8]{inputenc}

\usepackage{amssymb,amsmath,bm,bbold,graphicx}
\usepackage{ifpdf,color,mathrsfs}

\newcommand{\iu}{i} 
\newcommand{\de}{d} 
\newcommand{\ee}{e} 

\let\Re\undefined
\let\Im\undefined
\DeclareMathOperator{\Re}{Re}
\DeclareMathOperator{\Im}{Im}
\newcommand{\pdiff}[2]{\frac{\partial\,#1}{\partial #2}}

\newcommand{\bra}[1]{\ensuremath{\left\langle #1 \right|}}
\newcommand{\ket}[1]{\ensuremath{\left| #1\right\rangle}}

\newcommand{\melement}[3]{\ensuremath{\left\langle #1\left| #2 \right| #3\right\rangle}}

\begin{document}

\title{Ensemble properties of charge carriers injected by an ultrashort laser pulse}

\author{Muhammad Qasim}
\affiliation{Max-Planck-Institut f\"ur Quantenoptik, Hans-Kopfermann-Str. 1, Garching 85748, Germany}

\author{Michael S.~Wismer}
\affiliation{Max-Planck-Institut f\"ur Quantenoptik, Hans-Kopfermann-Str. 1, Garching 85748, Germany}

\author{Manoram Agarwal}
\affiliation{Max-Planck-Institut f\"ur Quantenoptik, Hans-Kopfermann-Str. 1, Garching 85748, Germany}

\author{Vladislav S.~Yakovlev}
\email[]{vladislav.yakovlev@mpq.mpg.de}
\affiliation{Max-Planck-Institut f\"ur Quantenoptik, Hans-Kopfermann-Str. 1, Garching 85748, Germany}
\affiliation{Ludwig-Maximilians-Universit\"at, Am Coulombwall~1, Garching 85748, Germany}

\date{\today}

\begin{abstract}
	The average effective mass of charge carriers produced by an intense ultrashort laser pulse in a transparent solid increases significantly as the excitation mechanism changes from multiphoton transitions to interband tunneling.
	We theoretically investigate this phenomenon for several dielectrics and semiconductors.
	For diamond as a representative dielectric, we present a detailed analysis of the laser-induced change of optical properties.
	When the concentration of free carriers is high, we find that the average effective mass controls not only the intraband charge-carrier transport but also the interband contributions to the optical response.
	We observe that the excitation-induced birefringence is particularly large for parameters where the plasma response compensates for the linear response of an unperturbed solid.
\end{abstract}


\maketitle

\section{Introduction}
The effective mass of charge carriers controls the optical and electric properties of solids.
When electrons and holes are produced by an intense laser pulse, their transient state is characterized by an average effective mass that may significantly exceed that in a state prepared by a weak laser pulse.
While this basic fact is well established~\cite{vanDriel_APL_1984, Yang_IEEE_1986, Sokolowski-Tinten_PRB_2000, Riffe_JOSAB_2002,Hoffmann_JOSAB_2009, Ulbricht_RMP_2011, Sato_PRB_2014_silicon, Sato_PRB_2014_temperature, Zhang_PRB_2018_hyperbolic_diamond} and explained by band nonparabolicity, the quest for extending the frontiers of ultrafast optoelectronic metrology requires a more detailed knowledge of the properties of photoexcited solids.
In this paper, we study the optical response of dielectrics and semiconductors excited by an intense few-cycle laser pulse, the spectrum of which lies within the medium's transparency region.
In this nonresonant regime, band nonparabolicity is essential when a laser pulse drives interband transitions within a large part of the first Brillouin zone.
This occurs when probabilities of various multiphoton excitation pathways become comparable to each other.
Consequently, band nonparabolicity is particularly important in the nonperturbative regime.

Early work on the optical effective mass of laser-excited carriers was motivated by the problem of optical determination of the carrier density~\cite{vanDriel_APL_1984}.
Since the Drude model operates with the density-to-mass ratio, the mass must be known to extract the density from reflection or transmission spectra.
Conversely, presuming the applicability of the Drude model, an optical measurement of the effective mass is possible only if the density of charge carriers is known.
Such measurements on silicon at the melting threshold showed a 20\% increase of the optical effective mass~\cite{Sokolowski-Tinten_PRB_2000}, confirming theoretical predictions~\cite{Yang_IEEE_1986}.
Hot free electrons and dense electron--hole plasmas were also investigated using terahertz pump--probe spectroscopy.
While band nonparabolicity played an important role in these experiments~\cite{Beard_PRB_2000, Blanchard_PRL_2011}, their research focus was on studying scattering phenomena.
In particular, it was observed that intervalley scattering leads to a significant change of the effective mass~\cite{Hoffmann_JOSAB_2009,Razzari_PRB_2009}, while deviations from the Drude model were explained by long-range transport and backscattering events~\cite{Cooke_PRB_2006, Ulbricht_RMP_2011, Shimakawa_APL_2012, Yang_IEEE_2013}.
Typical times for electron scattering lie between 100 femtoseconds and 100 picoseconds~\cite{Shah_1999,Sjakste_JPCM_2018}, even when the exciting laser field is strong~\cite{Cushing_SD_2018}.
Thus, to a first approximation, scattering is negligible during the interaction with a laser pulse that is as short as a few femtoseconds (unless the pulse is weak and indirect interband transitions dominate the optical response).
At the same time, the peak intensity of such a short pulse may be very high without inducing any damage, which allows one to study extremely nonlinear processes~\cite{Kruchinin_RMP_2018}.
Several recent theoretical papers report on the optical effective mass of charge carriers produced by such intense few-cycle laser pulses.
A good fit quality with the Drude model was reported in~[\onlinecite{Sato_PRB_2014_silicon}], where the average effective mass of charge carriers in silicon was predicted to increase, depending on the orientation, by 20--30\% upon the increase of the peak laser intensity from $10^{12}$~W/cm${}^2$ to $5 \times 10^{12}$~W/cm${}^2$.
The latter theoretical work employed the time-dependent density functional theory, and it was extended to finite electron temperatures~\cite{Sato_PRB_2014_temperature}, where the authors came to the following conclusion: ``In spite of the large difference of the electron--hole distributions between the thermal model and the numerical pump--probe simulation, the real parts of the dielectric functions are qualitatively similar.''
The time-dependent density functional theory was also applied to model how an intense ultrashort laser pulse changes the optical properties of diamond~\cite{Zhang_PRB_2017}, where one of the main findings was that the pulse may induce anisotropy in an isotropic solid.
Very recently, the same team predicted that, at extremely high intensities, the induced anisotropy reaches a level where laser-excited diamond may acquire a hyperbolic dispersion~\cite{Zhang_PRB_2018_hyperbolic_diamond}, where the real part of the permittivity is positive in one direction and negative in a perpendicular direction.
In the context of attosecond measurements, the formation of the effective mass after sudden excitation has recently been a matter of theoretical and experimental research~\cite{Fang_PRA_2014, Kasmi_Optica_2017}.

The purpose of this paper is to systematically analyze how an intense few-cycle laser pulse changes the optical properties of a transparent solid.
The motivation for this work came from several sources.
Apart from a lack of such an analysis in the literature, we wanted to point out that the average effective mass experiences a manifold increase within the parameter space relevant to ultrafast nondestructive measurements.
The dependence of the effective mass on laser-pulse parameters is important for measuring charge-carrier density~\cite{Temnov_PRL_2006, Mouskeftaras_APA_2013}, analyzing data acquired by ultrafast reflection~\cite{Sjodin_PRL_1998, Price_Nature-Communications_2015} and transmission~\cite{Winkler_APA_2006, Gertsvolf_JPB_2010, Sommer_Nature_2016, Popelar_Diamond_2017} spectroscopies in the strong-field regime, as well as interpreting time-resolved measurements of optical-field-driven charge-carrier transport~\cite{Schiffrin_Nature_2013, Kwon_SR_2016, Paasch-Colberg_Optica_2016}.

\section{Methods}

The main challenge in modeling the interaction of intense few-femtosecond laser pulses with solids is that interband transitions takes place in the entire Brillouin zone among many bands.
At the same time, the brief duration and the strength of the interaction allow one to make approximations that would be unjustified for longer, less intense laser pulses~\cite{Kruchinin_RMP_2018}.
From several recent experiments and their numerical analysis, we infer that relaxation processes, lattice motion, and electron--hole interaction usually play a minor role~\cite{Schultze_Science_2014,Lucchini_Science_2016,Schlaepfer_NaturePhysics_2018}.

For the purposes of this paper, we chose to solve the time-dependent Schr\"odinger equation (TDSE) in a stationary basis of Kohn--Sham orbitals, where the electron--electron interaction and correlation enter our model only by affecting band energies and transition matrix elements.
For diamond as a prototypical dielectric, this approximation has recently been shown to produce results that are very similar to those obtained with the time-dependent density-functional theory (TDDFT)~\cite{Floss_PRA_2018}, where the effect of electron--electron interaction was re-evaluated at every step of time propagation.
This is consistent with the observation that freezing the Coulomb and exchange-correlation terms in TDDFT calculations to their ground-state values tends to have a negligible effect on the polarization response of a bulk solid~\cite{Tancogne-Dejean_PRL_2016}.
In this case, local fields and band renormalization induced by exciting a small fraction of valence electrons can be neglected.
Under this presumption, it is advantageous to work in a stationary basis of Bloch states, rather than employ TDDFT.
The main advantage is flexibility.
Band energies, transition matrix elements and other input parameters can, in principle, be obtained with any suitable electronic-structure method: tight binding, density functional theory, quasiparticle self-consistent GW etc.

\subsection{Numerical simulations}
For each crystal momentum $\mathbf{k}$ and each initial valence band $n$, we solved the TDSE
\begin{equation} \label{eq:EOM}
\iu \hbar \dfrac{\de}{\de t} \ket{\psi_{n \mathbf{k}}(t)} =\left( \hat{H}_{\mathbf{k}}^{(0)} + \frac{e}{m_0}\mathbf{A}(t) \cdot \hat{\mathbf{p}} \right) \ket{\psi_{n \mathbf{k}}(t)}
\end{equation}
in the basis of stationary three-dimensional Bloch states $\ket{m \mathbf{k}}$:
\begin{equation} \label{eq:psi_ansatz}
	\ket{\psi_{n \mathbf{k}}(t)} = \sum_m \alpha_{m n}(\mathbf{k}, t)
	\ee^{-\frac{\iu}{\hbar} \epsilon_{m}(\mathbf{k}) t}
	\ket{m \mathbf{k}}.
\end{equation}
Here, $\hat{\mathbf{p}}$ is the momentum operator, $e>0$ is elementary charge, $m_0$ is the free-electron mass, $t_0$ is the starting time of a simulation, and the eigenstates of the unperturbed Hamiltonian are defined by $\hat{H}_{\mathbf{k}}^{(0)} \ket{m \mathbf{k}} = \epsilon_{m}(\mathbf{k}) \ket{m \mathbf{k}}$.
The expansion coefficients $\alpha_{m n}(\mathbf{k}, t)$ are the probability amplitudes of finding an electron in state $\ket{m \mathbf{k}}$ provided that the electron was initially in state $\ket{n \mathbf{k}}$.
So, the initial condition for solving Eq.~\eqref{eq:psi_ansatz} is $\alpha_{m n}(\mathbf{k}, t_0) = \delta_{m n}$.

In this velocity-gauge model, the electric field $\mathbf{F}(t)$ acting on electrons enters Eq.~\eqref{eq:EOM} via $\mathbf{A}(t)=-\int_{-\infty}^t \mathbf{F}(t')\,\de t'$.
We define the vector potential via
\begin{equation}
\mathbf{A}(t) = -\mathbf{e}_{\mathrm{L}} \frac{F_{\mathrm{L}}}{\omega_{\mathrm{L}}} \theta(T_{\mathrm{L}} - |t|)
\cos^4 \left( \frac{\pi}{2 T_{\mathrm{L}}} \right) \sin(\omega_{\mathrm{L}} t),
\end{equation}
where $\mathbf{e}_{\mathrm{L}}$ is a unit vector that defines the polarization of the laser pulse, $F_{\mathrm{L}}$ is approximately equal to the amplitude of the electric field, $\omega_{\mathrm{L}}$ is the pulse's central frequency, $\theta$ is the Heaviside step function, and $T_{\mathrm{L}} \ge -t_0 > 0$ is related to the full width at half maximum (FWHM) of the pulse intensity via $T_{\mathrm{L}} = \pi\,\mathrm{FWHM} / \left(4\,\text{arccos}(2^{-0.125})\right)$.

All the information about a medium that our numerical model requires is $\epsilon_{m}(\mathbf{k})$ and the matrix elements of the momentum operator:
\begin{equation}
	\mathbf{p}_{m n}(\mathbf{k}) = \melement{m \mathbf{k}}{\hat{\mathbf{p}}}{n \mathbf{k}},
\end{equation}
where the integration is performed over a unit cell.
We obtained this input data from density functional theory using standard packages: Wien2k~\cite{Schwarz_CPC_2002} for SiO${}_2$ and Abinit~\cite{Gonze_CPC_2009} for all the other solids.
For most of our calculations, we used the Tran--Blaha correction to the Becke--Johnson meta-GGA exchange--correlation potential with Perdew--Wang correlation.
This functional is known to produce more accurate values of the energy band gap as compared to the local density approximation (LDA)~\cite{Tran_PRL_2009}.
The energy cutoff was set to 19 Hartree, and we used a nonshifted Monkhorst-Pack $\mathbf{k}$ grid.

Most of the results in this paper were obtained by analyzing occupations at the end of the laser pulse:
\begin{equation}
	f_m(\mathbf{k}) = \sum_{n \in \text{VB}} |\alpha_{m n}(\mathbf{k}, T_{\mathrm{L}})|^2,
\end{equation}
where we add contributions from all the valence bands (VB).
In Sec.~\ref{sec:pump-probe}, we also show results that require the evaluation of the electric current density, $\mathbf{J}(t)$.
It is convenient to express $\mathbf{J}(t)$ via the density operator:
\begin{align}
	\hat \rho_{\mathbf{k}}(t) &= \sum_{n \in \text{VB}}
	\ket{\psi_{n\mathbf{k}}(t)} \bra{\psi_{n\mathbf{k}}(t)},\\
	\label{eq:current}
	\mathbf{J}(t) &=  - \frac{e}{m_0}
	\int_{\mathrm{BZ}}\! \frac{\de^3 \mathbf{k}}{\left(2\pi \right)^3}\,  \text{Tr} \Bigl[ \hat{\rho}_{\mathbf{k}}(t)  \bigl(\hat{\mathbf{p}} + e \mathbf{A}(t) \bigr) \Bigr] + \Delta\mathbf{J}(t).
\end{align}
Here, the integral is taken over the first Brillouin zone (BZ), and $\Delta\mathbf{J}(t)$ is an adiabatic correction introduced in~[\onlinecite{Yakovlev_CPC_2017}].
Note that Eq.~\eqref{eq:current} does not explicitly account for spin degeneracy (if the valence states are doubly occupied, the right-hand side of this equation needs to be multiplied with 2).

\subsection{Linear response}
Let us consider a solid excited by an intense laser pulse.
Investigating the laser-induced change of a medium's optical properties, we are interested in its response to a weak probe pulse, the electric field of which is $\mathbf{F}_{\mathrm{probe}}(t) = -\mathbf{A}_{\mathrm{probe}}'(t)$.
This response can be calculated numerically, using the method described in the previous subsection, or analytically, using the standard time-dependent perturbation theory.
The analytical approach has two main advantages: it allows us to decompose the laser-induced change of optical properties into intra- and interband components, and it also obviates the necessity to control numerical convergence when we consider the limit of an infinitesimally weak probe pulse.
However, the expressions derived below are applicable only to non-overlapping pump and probe pulses.

The first-order perturbation theory yields the following expression for the electric current density induced by a weak probe pulse:
\begin{multline} \label{eq:J_probe_t}
	\mathbf{J}_{\mathrm{probe}}(t) =\\ -\frac{e^2}{m_0}
	\int_{\mathrm{BZ}} \frac{\de^3 \mathbf{k}}{\left(2\pi \right)^3}\, 
	\Biggl\{ \sum_n f_n(\mathbf{k}) \Biggl[ \frac{\bigl( \mathbf{A}_{\mathrm{probe}}(t) \nabla_\mathbf{k} \bigr) \mathbf{p}_{n n}(\mathbf{k})}{\hbar} \\+
	\frac{2}{m_0} \sum_{m \ne n} \frac{\Re\left[\bigl( \mathbf{p}_{n m}(\mathbf{k}) \mathbf{A}_{\mathrm{probe}}(t) \bigr) \mathbf{p}_{m n}(\mathbf{k})\right]}{\hbar \omega_{m n}} \Biggr] \\+
	\int_0^{\infty} \de \tau\,
	\hat{s}(\mathbf{k}, \tau, t) \mathbf{A}_{\mathrm{probe}}(t-\tau) \Biggr\},
\end{multline}
where the Cartesian components of the $\hat{s}$ tensor are given by
\begin{multline} \label{eq:sigma}
	s_{\alpha \beta}(\mathbf{k}, \tau, t) = \frac{2}{\hbar m_0}
	\Im\biggl[
		\sum_{n m} 
		\ee^{-\iu (t - T_{\mathrm{L}} - \tau) \omega_{n m}(\mathbf{k})}\\
		\times \rho_{n m}(\mathbf{k}) \sum_{n'} p_{m n'}^{\alpha}(\mathbf{k}) p_{n' n}^{\beta}(\mathbf{k})
		\ee^{-\iu \tau \omega_{n' m}(\mathbf{k})}
	\biggr].
\end{multline}
Here,
\begin{equation}
	\rho_{n m}(\mathbf{k}) = \melement{n \mathbf{k}}{\hat \rho_{\mathbf{k}}(T_\mathrm{L})}{m \mathbf{k}}
\end{equation}
is the density matrix at the end of the pump pulse [$\rho_{n n}(\mathbf{k}) \equiv f_n(\mathbf{k})$],
$p_{m n}^{\alpha}(\mathbf{k}) = \mathbf{e}_{\alpha} \mathbf{p}_{m n}(\mathbf{k})$ denotes the Cartesian components of the momentum matrix element ($\alpha \in \{x,y,z\}$, $\mathbf{e}_{\alpha}$ is a unit vector), and we have introduced the transition frequencies:
\begin{equation}
	\omega_{m n}(\mathbf{k}) = \frac{\epsilon_{m}(\mathbf{k}) - \epsilon_{n}(\mathbf{k})}{\hbar}.
\end{equation}
Equation \eqref{eq:J_probe_t} incorporates first-order adiabatic velocity-gauge corrections~\cite{Yakovlev_CPC_2017}.
These corrections compensate for numerical artifacts arising in the velocity gauge due to the violation of the Thomas-Reiche-Kuhn rule caused, e.g., by basis truncation.

The model defined by Eqs.~\eqref{eq:J_probe_t} and \eqref{eq:sigma} considerably simplifies if one neglects interband coherences, that is, the off-diagonal elements of $\rho_{n m}(\mathbf{k})$.
In the next section, we provide some evidence that this is a reasonable approximation; we also verified this approximation by directly by evaluating Eq.~\eqref{eq:sigma} with and without interband coherences.
Once the off-diagonal elements of $\rho_{n m}(\mathbf{k})$ are neglected, $s_{\alpha \beta}(\mathbf{k}, \tau, t)$ no longer depends on $t$, so that the integration over $\tau$ in Eq.~\eqref{eq:J_probe_t} becomes a convolution.
In this case, the linear response of the medium can be described with the tensor of linear susceptibility, $\hat{\chi}^{(1)} $, which we first define in the time domain. In CGS units,
\begin{equation} \label{eq:chi_TD}
	\mathbf{P}_{\mathrm{probe}}(t) = \int_0^t \de\tau\,
	\hat{\chi}^{(1)}(\tau) \mathbf{F}_{\mathrm{probe}}(t - \tau).
\end{equation}
Let us distinguish between the intra- and interband contributions deriving $\hat{\chi}^{(1)}(\tau)$ from Eqs.~\eqref{eq:J_probe_t} and \eqref{eq:sigma}:
\[
	\hat{\chi}^{(1)} = \hat{\chi}^{\mathrm{intra}} + \hat{\chi}^{\mathrm{inter}}.
\]
The terms that enter $\hat{\chi}^{\mathrm{intra}}$ must not contain matrix elements describing transitions between different states, while $\hat{\chi}^{\mathrm{inter}}$ may only contain off-diagonal elements of the momentum matrix.
Neglecting interband coherences, we obtain
\begin{equation}
	\hat{\chi}^{\mathrm{intra}}(\tau) = \theta(\tau) \frac{e^2 \tau}{\hbar m_0}
	\int_{\mathrm{BZ}} \frac{\de^3 \mathbf{k}}{\left(2\pi \right)^3}\, 
	\sum_n f_n(\mathbf{k}) \pdiff{p_{n n}^\alpha}{k_{\beta}}
\end{equation}
and
\begin{multline}
\chi_{\alpha \beta}^{\mathrm{inter}}(\tau) = \ee^{-\gamma \tau} \theta(\tau) \frac{2 e^2}{\hbar m_0^2}
\int_{\mathrm{BZ}} \frac{\de^3 \mathbf{k}}{\left(2\pi \right)^3}\, 
\sum_n \biggl\{ f_n(\mathbf{k}) \\ \times
\Im\biggl[
\sum_{m \ne n} p_{n m}^{\alpha}(\mathbf{k}) p_{m n}^{\beta}(\mathbf{k})
\frac{1 - \ee^{-\iu \tau \omega_{m n}(\mathbf{k})}}{\omega_{m n}^2(\mathbf{k})}
\biggr] \biggr\},
\end{multline}
where $\theta(\tau)$ is the Heaviside step function, and we have introduced a phenomenological decoherence rate, $\gamma = T_2^{-1}$.
Without decoherence, the absorption spectrum in the numerical model would consist of a discrete set of infinitely narrow absorption lines.

In the frequency domain,
\begin{equation}
	\mathbf{P}_{\mathrm{probe}}(\omega) =
	\hat{\chi}^{(1)}(\omega) \mathbf{F}_{\mathrm{probe}}(\omega).
\end{equation}
Using $\mathcal{F}[f(t)] = \int_{-\infty}^{\infty} f(t) \exp[\iu \omega t] dt$ as the definition of the Fourier transform and employing the well-known expression for the inverse-mass tensor,
\begin{equation} \label{eq:inverse_mass_theorem}
\left(\hat{m}^{-1}(n, \mathbf{k}) \right)_{\alpha \beta} =
\frac{1}{\hbar m_0} \pdiff{p_{n n}^\alpha}{k_{\beta}},
\end{equation}
we arrive at the following expressions for the frequency-domain intraband tensors of linear susceptibility:
\begin{equation} \label{eq:chi_intra}
\hat{\chi}^{\mathrm{intra}}(\omega) =
-\frac{e^2}{\omega^2}
\int_{\mathrm{BZ}} \frac{\de^3 \mathbf{k}}{\left(2\pi \right)^3}\, 
\sum_n f_n(\mathbf{k}) \hat{m}^{-1}(n, \mathbf{k}),
\end{equation}
\begin{multline} \label{eq:chi_inter}
\chi_{\alpha \beta}^{\mathrm{inter}}(\omega) = \frac{e^2}{\hbar m_0^2}
\int_{\mathrm{BZ}} \frac{\de^3 \mathbf{k}}{\left(2\pi \right)^3}\, 
\sum_n \biggl\{ f_n(\mathbf{k}) \\
\times
\sum_{m \ne n} \biggl[ \frac{1}{\omega_{m n}^2(\mathbf{k})} \biggl(
  \frac{2 \iu}{\omega + \iu \gamma}
  \Im\left[ p_{n m}^{\alpha}(\mathbf{k}) p_{m n}^{\beta}(\mathbf{k}) \right] \\+
  \frac{p_{n m}^{\alpha}(\mathbf{k}) p_{m n}^{\beta}(\mathbf{k})}
    {\omega_{m n}(\mathbf{k}) - \omega - \iu \gamma} +
  \frac{p_{m n}^{\alpha}(\mathbf{k}) p_{n m}^{\beta}(\mathbf{k})}
    {\omega_{m n}(\mathbf{k}) + \omega + \iu \gamma}
\biggr) \biggr] \biggr\}.
\end{multline}

Let us compare the intraband susceptibility to that in the collisionless Drude model:
\begin{equation*}
	\hat{\chi}^{\mathrm{Drude}}(\omega) =
	-\frac{e^2}{\omega^2}
	\left( N_e \hat{m}_{e}^{-1} + N_h \hat{m}_{h}^{-1} \right),
\end{equation*}
where $N_e$ and $N_h$ are the concentrations of electrons and holes, while $\hat{m}_{e}^{-1}$ and $\hat{m}_{h}^{-1}$ are their average inverse-mass tensors.
Assuming that there are no charge carries in the ground state (before the pump pulse), we set $N_h = N_e = N_{e-h}$, define the average tensor of the reduced inverse mass as $\hat{m}_{\mathrm{eff}}^{-1} = \hat{m}_{e}^{-1} + \hat{m}_{h}^{-1}$, and write
\begin{equation} \label{eq:chi_Drude}
	\hat{\chi}^{\mathrm{Drude}}(\omega) =
	-\frac{e^2}{\omega^2} N_{e-h} \hat{m}_{\mathrm{eff}}^{-1}.
\end{equation}
Henceforth, we will refer to $\hat{m}_{\mathrm{eff}}^{-1}$ as the average inverse mass.
Comparing Eq.~\eqref{eq:chi_Drude} with Eq.~\eqref{eq:chi_intra} and using
\begin{equation*}
	N_{e-h} = \int_{\mathrm{BZ}} \frac{\de^3 \mathbf{k}}{\left(2\pi \right)^3}\, 
	\sum_{n \in \mathrm{CB}} f_n(\mathbf{k}),
\end{equation*}
where the summation is performed over conduction bands (CB), we define the average inverse mass as
\begin{equation} \label{eq:average_inverse_mass}
	\hat{m}_{\mathrm{eff}}^{-1} = 
	\frac{
		\int_{\mathrm{BZ}} \de^3 \mathbf{k}\, \sum_n f_n(\mathbf{k}) \hat{m}^{-1}(n, \mathbf{k})
	}{ \int_{\mathrm{BZ}} \de^3 \mathbf{k}\, \sum_{n \in \mathrm{CB}} f_n(\mathbf{k}) }.
\end{equation}
Not surprisingly, this result represents averaging the inverse-mass tensor over the ensemble of electrons and holes using the occupation numbers as averaging weights.

In the next section, we investigate the optical response to a linearly polarized laser pulse.
Let $\mathbf{e}_{\mathrm{probe}}$ be a unit vector that is parallel to the electric field of the probe pulse.
We evaluate the average inverse mass with respect to the probe pulse as
\begin{equation} \label{eq:scalar_inverse_mass}
	m_{\mathrm{eff}}^{-1} = \mathbf{e}_{\mathrm{probe}} \left( \hat{m}_{\mathrm{eff}}^{-1} \mathbf{e}_{\mathrm{probe}} \right).
\end{equation}
The average effective mass with respect to this pulse is then defined by $m_{\mathrm{eff}} = 1 / m_{\mathrm{eff}}^{-1}$.

\section{Results and discussion}


\subsection{Comparison of several crystals}
\begin{figure}
	\begin{tabular}{p{0.95\columnwidth} p{0pt}}
		\vspace{0mm} \includegraphics[width=0.9\columnwidth]{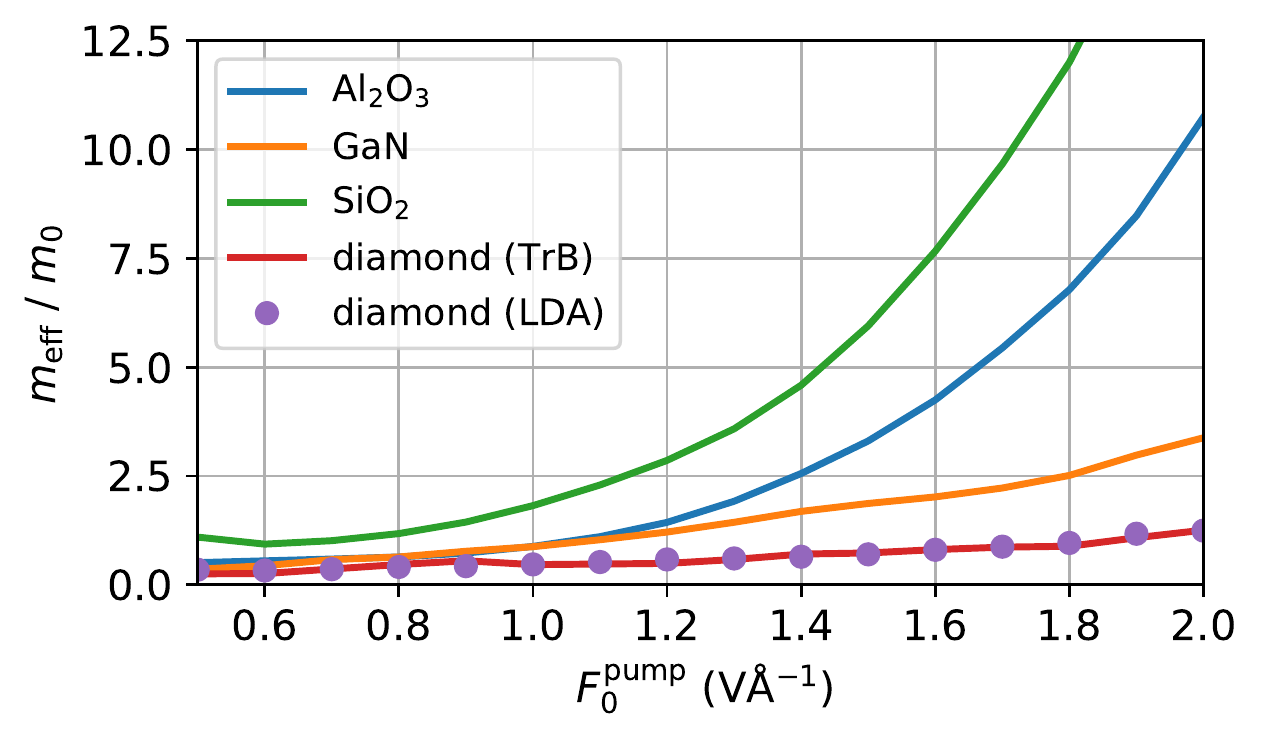} &
		\vspace{1mm} \hspace{-\columnwidth}
		\textbf{(a)}
	\end{tabular}\\[-4mm]
	\begin{tabular}{p{0.95\columnwidth} p{0pt}}
		\vspace{0mm} \includegraphics[width=0.9\columnwidth]{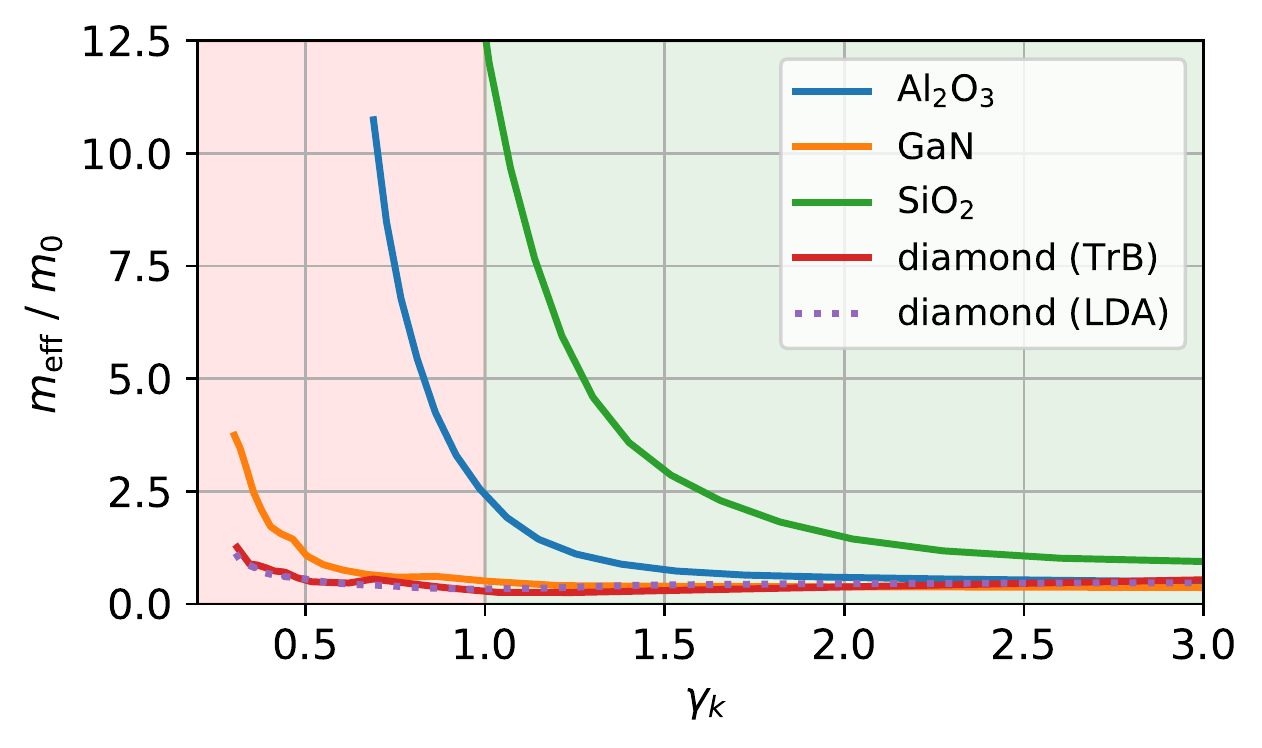} &
		\vspace{1mm} \hspace{-\columnwidth}
		\textbf{(b)}
	\end{tabular}
	\caption{\label{fig:overview}
		(a) The dependence of the average effective mass on the peak electric field of a 4-fs 800-nm laser pulse.
		For diamond, we compare data obtained with two exchange--correlation potentials: Tran--Blaha~\cite{Tran_PRL_2009} (TrB) and local-density approximation (LDA).
		For all the other solids, we show only the outcomes of calculations with the Tran--Blaha functional.
		(b) The same data as in panel (a) plotted against the Keldysh parameter.
	}
\end{figure}

In Fig.~\ref{fig:overview}, we compare average effective masses \eqref{eq:average_inverse_mass}	 for several solids excited by a 4-fs 800-nm laser pulse.
The optic axis of each uniaxial crystal was taken as the laser-beam axis.
For Al${}_2$O${}_3$, SiO${}_2$, and GaN, the optic axis was parallel to $[001]$, and we took the $[100]$ direction as the polarization direction of the pump pulse, $\mathbf{e}_{\mathrm{pump}}$.
For diamond, which is an isotropic medium, we took $[111]$ as the beam axis and $\mathbf{e}_{\mathrm{pump}} \parallel [1\bar{1}0]$.
We evaluated the effective masses using Eqs.~\eqref{eq:average_inverse_mass} and \eqref{eq:scalar_inverse_mass} with $\mathbf{e}_{\mathrm{probe}}$ being perpendicular to both $\mathbf{e}_{\mathrm{pump}}$ and the optic axis.

The peak electric field in this and other figures is the field within the medium.
The relationship of this field with the vacuum field of an incident laser pulse may be complex because laser-induced interband transitions change the reflectivity and may be responsible for complex propagation effects.
Since our goal is to study the general properties of electron--hole plasmas created by an intense few-cycle laser pulse, we assume that the laser pulse in our simulations represents a pulse that has propagated to a particular position within a solid.
The maximal electric field that we used for Fig.~\ref{fig:overview} was 2~V/{\AA}, which is close to the damage threshold of the considered materials. 
At 1~V/{\AA}, we do not expect any of these materials to be damaged by such a laser pulse (at this field strength, we observe the highest concentration of excited electrons in GaN, where it takes a value of $1.9 \times 10^{21}$ $\mathrm{cm}^{-3}$).

Figure~\ref{fig:overview} thus illustrates that a significant increase of the effective mass is a general effect observed in many solids.
As we mentioned in the Introduction, this increase is expected upon the transition from the multiphoton excitation regime to the tunneling one because the multiphoton regime presumes that the probability of absorbing $N$ photons rapidly decreases with $N$ (as long as the transitions are energetically allowed).
If the probability of absorbing $N+1$ photons is much smaller than that of absorbing $N$ photons, then a laser pulse will usually inject carriers within a fraction of the Brillouin zone, where the lowest multiphoton order dominates.
As long as the involved valence and conduction bands are approximately parabolic within this reciprocal-space volume, their contributions to the average effective mass are approximately  independent of the peak injection field.
To support these arguments with evidence, we plot in Fig.~\ref{fig:overview}(b) the average effective masses as functions of the Keldysh parameter~\cite{Keldysh_JETP_1965}
\begin{equation}
	\gamma_{K} = \frac{\omega_{\mathrm{L}}}{|e F_0^{\mathrm{pump}}|} \sqrt{E_{g} m^*},
\end{equation}
where $E_{g}$ is the direct band gap.
Here, we evaluated the reduced mass, $m^*$, at the $\Gamma$ point.
Since some bands are degenerate at $\mathbf{k} = 0$, we did not limit the band selection to the top valence and bottom conduction bands.
Instead, we calculated a weighted average over all the bands: for each band, we calculated its curvature in the direction of the pump field and multiplied it with the concentration of charge carriers created in this band by the 1-V/{\AA} pulse.
The Keldysh parameter classifies the regimes of interband transitions into multiphoton ($\gamma_{K} \gg 1$) and tunneling ($\gamma_{K} \ll 1$).
From Fig.~\ref{fig:overview}(b) we see that the transition from the multiphoton regime to the tunneling one is indeed accompanied by a large increase of the average effective mass.

\subsection{The composition of the average effective mass}
In this subsection, we examine various factors that contribute to the average effective mass and its properties.
As a representative crystal for our analysis, we chose diamond---a medium with isotropic linear properties, a relatively simple band structure, and a high damage threshold.
Figure \ref{fig:overview} shows that, for diamond, the Tran--Blaha and LDA exchange--correlation potentials produce similar average effective masses.
We note, however, that there are considerable differences in the electronic structure.
For example, the direct band gaps in the TrB and LDA calculations were 6.4~eV and 5.5~eV, respectively.
Even though the LDA exchange--correlation potential is known to underestimate band gaps, it is a well-studied approximation where the effective potential experienced by each electron is local, which gives us more confidence in our numerical results (in general, nonlocal potentials lead to additional terms in velocity-gauge propagation equations).
Therefore, we use the LDA for the numerical analysis in this and following subsections.

\begin{figure}
	\includegraphics[width=0.9\columnwidth]{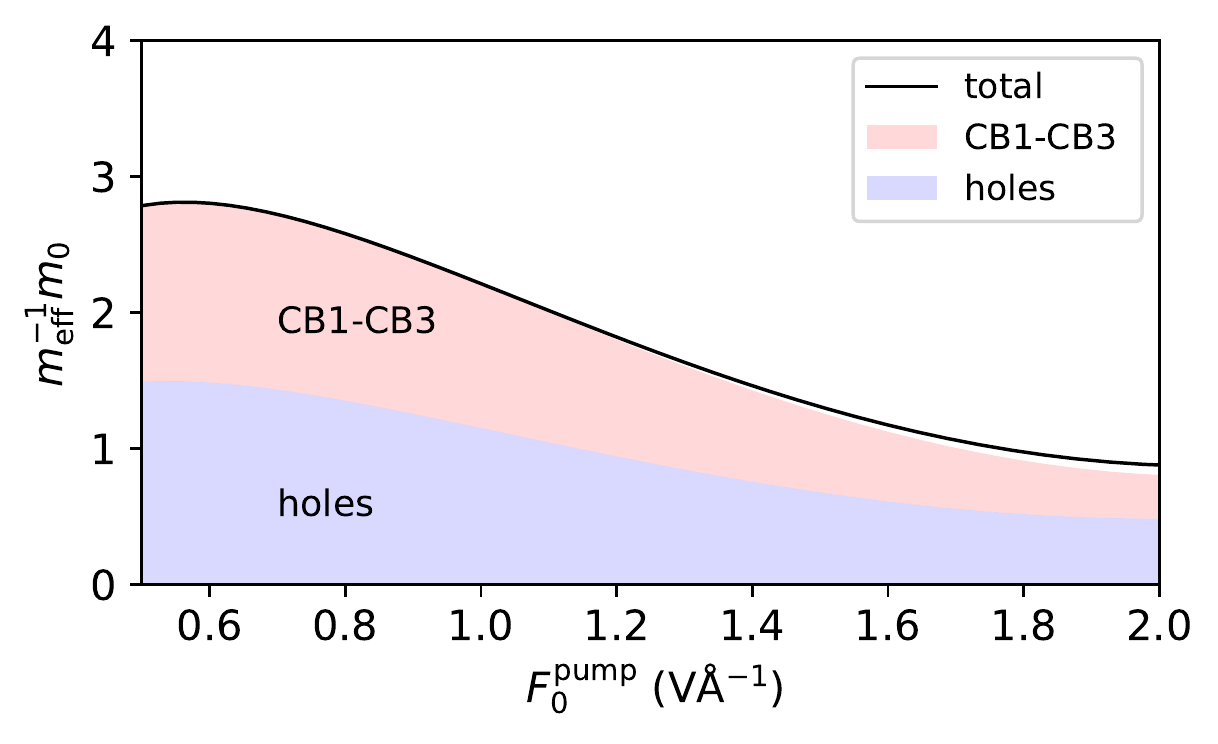}
	\caption{\label{fig:electrons_vs_holes}
		Contributions to the average inverse mass of charge carriers (solid black curve) from the valence-band states (holes), as well as from the lowest three conduction bands (CB1-CB3) in diamond.
	}
\end{figure}

In Fig.~\ref{fig:electrons_vs_holes}, we show how different bands contribute to the average inverse mass.
The lowest three conduction bands of diamond are degenerate at the $\Gamma$ point (neglecting the spin--orbit interaction), so we plot the sum of their contributions.
We also plot the inverse mass averaged over all the holes, as well as the net inverse mass obtained by averaging over all the charge carriers.
Electrons and holes in diamond have comparable average masses, which is why valence and conduction bands make comparable contributions to $m_{\mathrm{eff}}^{-1}$.
We also see that higher conduction bands contribute surprisingly little.
After the 2-V/{\AA} pulse, 28.1\% of excited electrons reside in bands above the third conduction band, but their relative contribution to the average inverse mass is as little as 9.4\%.

\begin{figure}
	\begin{tabular}{p{0.46\columnwidth} p{0pt}}
		\vspace{0mm} \includegraphics[width=0.46\columnwidth]{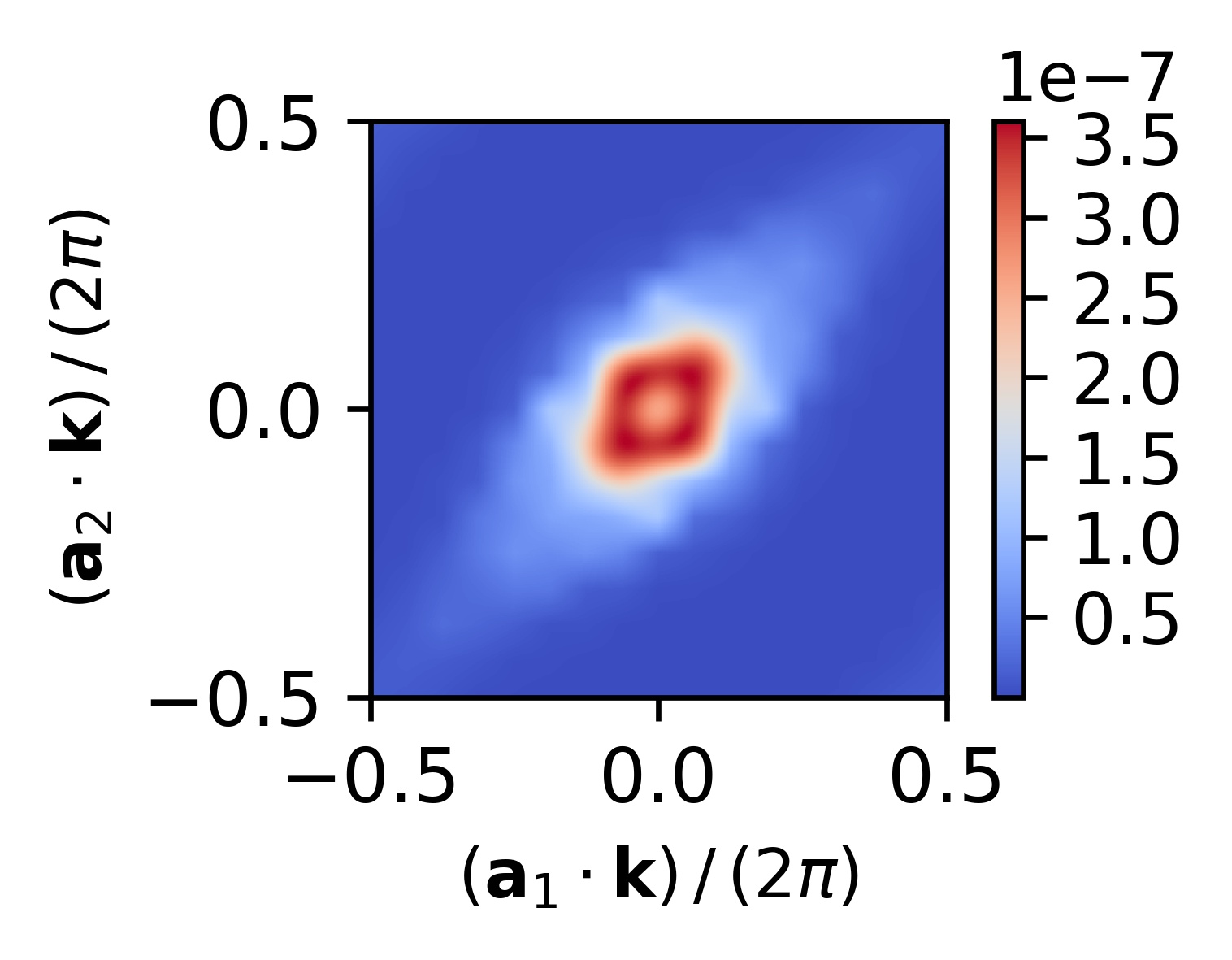} &
		\vspace{2mm} \hspace{-0.51\columnwidth}
		\textbf{(a)}
	\end{tabular}
	\begin{tabular}{p{0.46\columnwidth} p{0pt}}
		\vspace{0mm} \includegraphics[width=0.46\columnwidth]{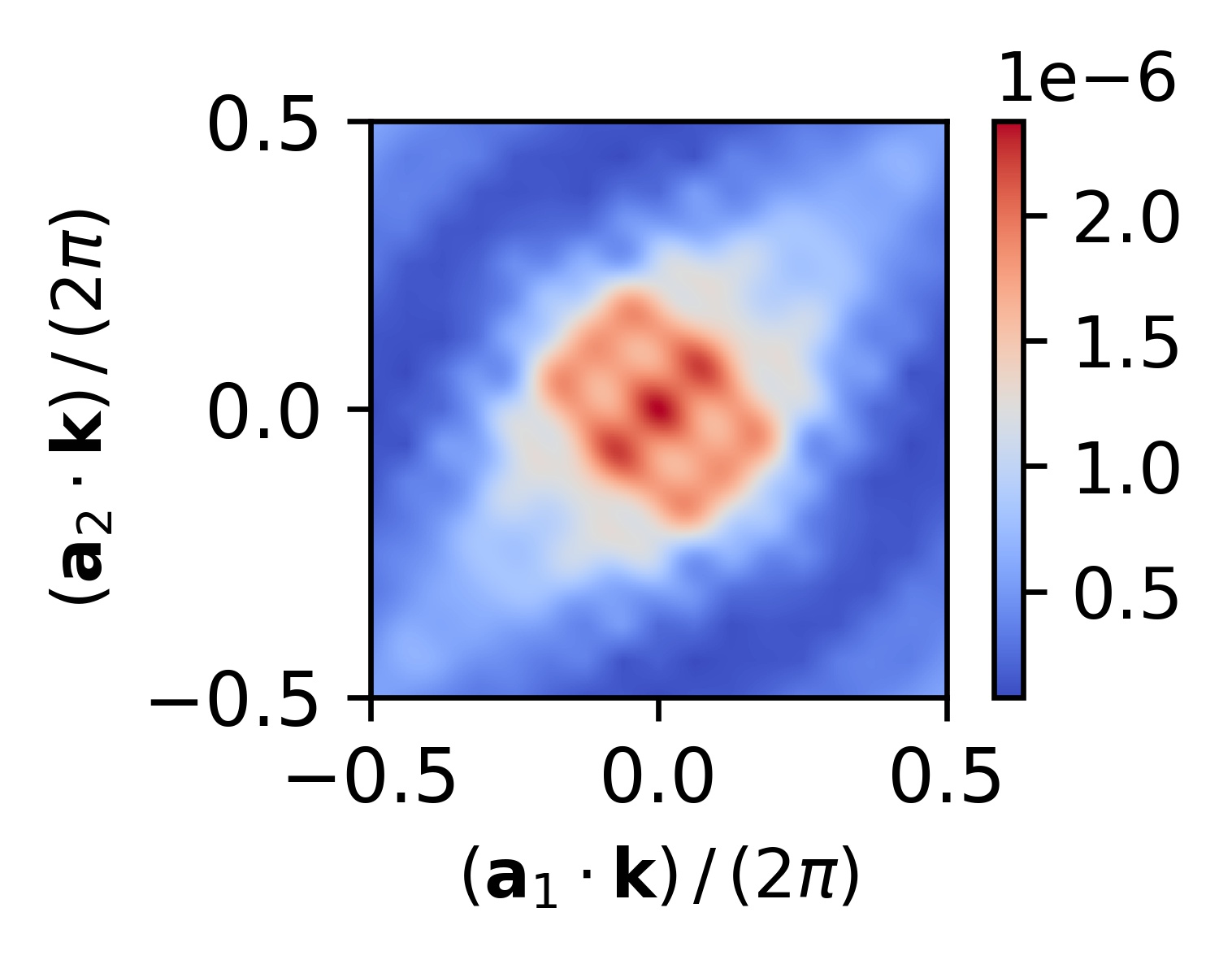} &
		\vspace{2mm} \hspace{-0.51\columnwidth}
		\textbf{(b)}
	\end{tabular}\\[-3mm]
	\begin{tabular}{p{0.46\columnwidth} p{0pt}}
		\vspace{0mm} \includegraphics[width=0.46\columnwidth]{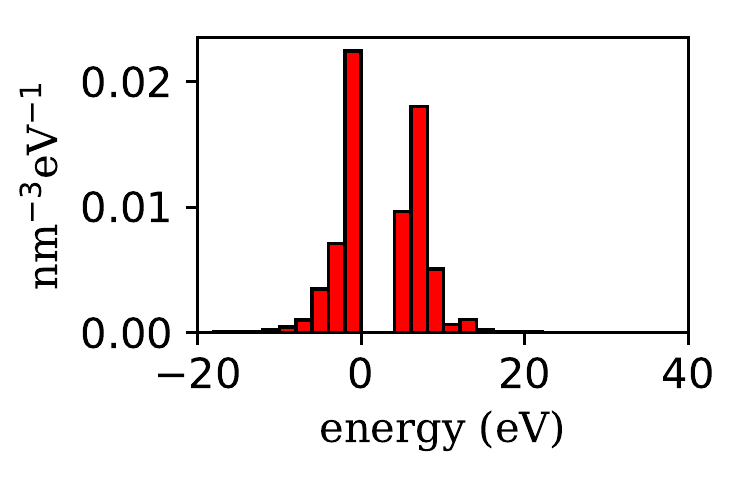} &
		\vspace{-1mm} \hspace{-0.51\columnwidth}
		\textbf{(c)}
	\end{tabular}
	\begin{tabular}{p{0.46\columnwidth} p{0pt}}
		\vspace{0mm} \includegraphics[width=0.46\columnwidth]{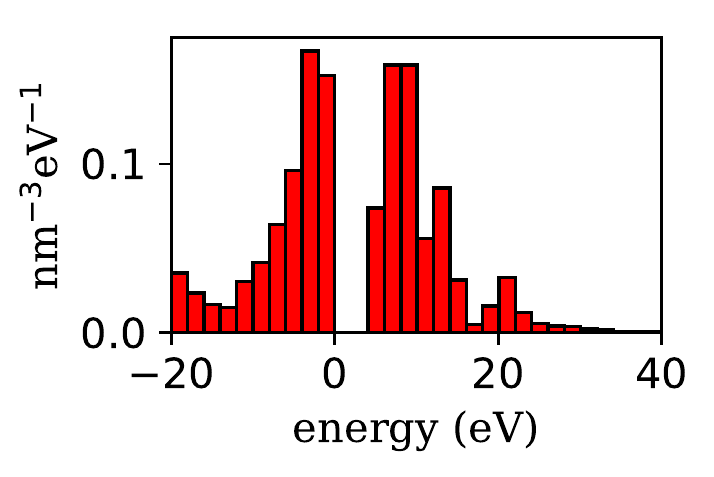} &
		\vspace{-1mm} \hspace{-0.51\columnwidth}
		\textbf{(d)}
	\end{tabular}\\[-3mm]
	\begin{tabular}{p{0.46\columnwidth} p{0pt}}
		\vspace{0mm} \includegraphics[width=0.46\columnwidth]{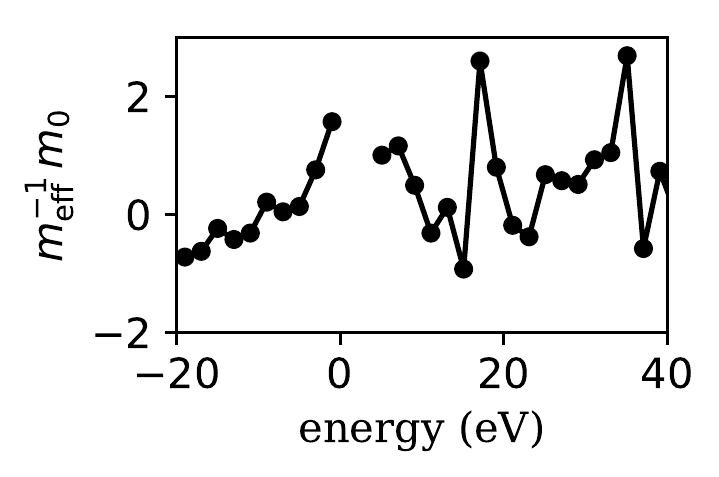} &
		\vspace{0mm} \hspace{-0.51\columnwidth}
		\textbf{(e)}
	\end{tabular}
	\begin{tabular}{p{0.46\columnwidth} p{0pt}}
		\vspace{0mm} \includegraphics[width=0.46\columnwidth]{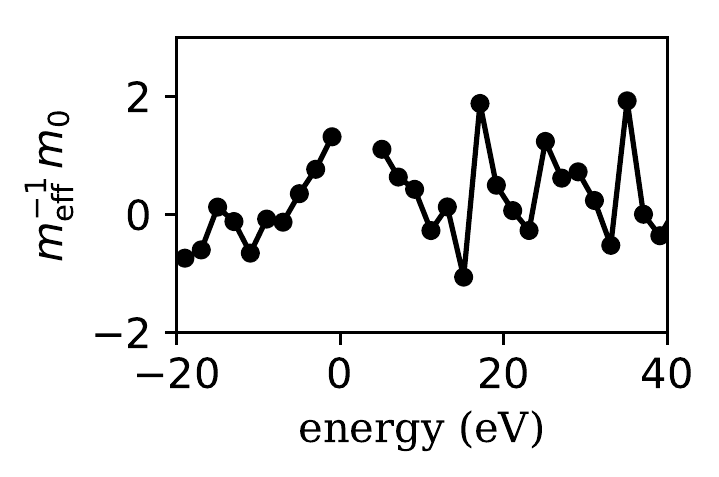} &
		\vspace{0mm} \hspace{-0.51\columnwidth}
		\textbf{(f)}
	\end{tabular}
	\caption{\label{fig:mass_explanation}
		The left and right panels represent simulations with peak laser fields of 1~V/{\AA} and 2~V/{\AA}, respectively.
		(a,b) The probability density of exciting a valence-band electron at a certain crystal momentum.
		The probability densities were integrated over the first Brillouin zone along the $\mathbf{b}_3$ vector.
		(c,d) The area of each bar represents the number of charge carriers with an energy within the corresponding 2-eV-wide bin.
		(e,f) The average inverse masses of charge carriers (holes for negative energies and electrons for positive ones) within the energy bins that were used for (c) and (d).
	}
\end{figure}

Figure \ref{fig:mass_explanation} gives a more detailed view on what contributes to the dependence of the effective mass on the peak laser field.
For the 1- and 2-V/{\AA} pulses, we plot $\mathbf{k}$- and energy-dependent quantities that determine the average mass of charge carriers.
Panels (a) and (b) of this figure visualize excitations within the primitive unit cell in reciprocal space, where we added the diagonal elements of the density matrix for all the conduction bands and integrated the transition probabilities along the $\mathbf{b}_3$ vector.
More precisely, these two false-color diagrams visualize
\begin{align}
 y(\xi_1, \xi_2) &= \sum_{n \in \mathrm{CB}}
\int_{-1/2}^{1/2} \de\xi_3 \rho_{nn}\left(\mathbf{k} \right),\\
 \mathbf{k} &= \xi_1 \mathbf{b}_1 + \xi_2 \mathbf{b}_2 + \xi_3 \mathbf{b}_3,
\end{align}
where $\mathbf{b}_i$ are the primitive vectors of the reciprocal lattice.
In their basis, the coordinates of the crystal momentum are $\xi_i = (\mathbf{a}_i \mathbf{k}) / (2 \pi)$, where $\mathbf{a}_i$ are Bravais lattice vectors.
Figures \ref{fig:mass_explanation}(a) and  \ref{fig:mass_explanation}(b) illustrate that an intense laser pulse drives transitions within a substantial part of the first Brillouin zone.

The rigorous definition of the average effective mass demands the knowledge of $\rho_{nn}(\mathbf{k})$.
Nevertheless, in our experience, 
the shape of a reciprocal-space excitation pattern is insignificant.
We illustrate it in Figs.~\ref{fig:mass_explanation}(c)-(f).
In Figs.~\ref{fig:mass_explanation}(c) and \ref{fig:mass_explanation}(d), each bar represents the concentration of charge carriers in states where the energy belongs to the corresponding 2-eV-broad energy bin.
These probability distributions are very sensitive to $F_0^{\mathrm{pump}}$.
For each bin, we also calculate the average inverse mass of charge carriers within the bin's energy range and plot the result in Figs.~\ref{fig:mass_explanation}(e) and \ref{fig:mass_explanation}(f).
These energy-dependent inverse masses depend on how the $\ket{\mathbf{k},n}$ states within a particular energy bin are populated, that is, on the reciprocal-space excitation pattern.
However, this dependence is weak, as we see by comparing Figs.~\ref{fig:mass_explanation}(e) and \ref{fig:mass_explanation}(f).
Knowing such an energy-dependent effective mass for a representative laser pulse, one can directly relate energy-dependent occupation numbers to the average effective mass.
For example, using the 2-V/{\AA} occupations [Fig.~\ref{fig:mass_explanation}(d)] to average the 1-V/{\AA} inverse masses from Fig.~\ref{fig:mass_explanation}(e) yields an average inverse mass of $1.0 m_0$, while the accurate value from Eq.~\eqref{eq:average_inverse_mass} is $0.8 m_0$.
Applying the same procedure to extrapolate from 1~V/{\AA} to 0.5~V/{\AA}, we get $2.6 m_0$ as the estimation of inverse mass, which is close to the accurate value of $2.9 m_0$ at 0.5~V/{\AA}.

\subsection{Intra- and interband contributions to the optical response}
When a laser pulse excites electrons from valence to conduction bands, the linear susceptibility of the solid, $\hat{\chi}^{(1)}(\omega)$, changes.
We decompose this change, $\Delta \hat{\chi}^{(1)}(\omega)$, into two components:
We evaluate the intraband component, $\Delta \hat{\chi}^{\mathrm{intra}}(\omega)$, assuming that the interaction with a weak probe pulse consists in changing the crystal momentum of each charger carrier according to the acceleration theorem, disregarding transitions between bands.
This is equivalent to the Drude model that neglects relaxation (scattering) processes.
In this model, $\omega^2 \Delta \hat{\chi}^{\mathrm{intra}}(\omega)$ is a frequency-independent real-valued quantity (see Eq.~\eqref{eq:chi_intra}) representing the response to an infinitesimally weak long-wavelength probe pulse.

The difference between the total and intraband changes of the linear susceptibility is the {interband component:
\begin{equation}
\Delta \hat{\chi}^{\mathrm{inter}}(\omega) = \Delta \hat{\chi}^{(1)}(\omega) - \Delta \hat{\chi}^{\mathrm{intra}}(\omega),
\end{equation}
which represents the optical response due to transitions between bands.
Even if a probe pulse has no frequency components above the band edge, it is responsible for two types of interband dynamics: virtual transitions describe a transient polarization induced by the probe pulse, while real transitions among valence or conduction bands describe excitation and de-excitation of charge carriers left by a pump pulse.
If the frequency of a probe field exceeds the band edge, then $\Delta \hat{\chi}^{\mathrm{inter}}(\omega)$ also reflects transitions between valence- and conduction-band states driven by a weak probe pulse.

In Fig.~\ref{fig:intra_vs_total_spectrum}, we compare the real part of
\[ \Delta \chi^{(1)}(\omega) \equiv
	\mathbf{e}_{\mathrm{probe}} \bigl(\Delta \hat{\chi}^{(1)}(\omega) \mathbf{e}_{\mathrm{probe}}\bigr)
\]
with
\[ \Delta \chi^{\mathrm{intra}}(\omega) \equiv
	\mathbf{e}_{\mathrm{probe}} \bigl(\Delta \hat{\chi}^{\mathrm{intra}}(\omega) \mathbf{e}_{\mathrm{probe}}\bigr)
\]
for a 1-V/{\AA} pulse interacting with diamond in the local-density approximation.
Below the band edge, the electron--hole plasma created by the laser pulse decreases the real part of the linear susceptibility.
The intraband contribution to the susceptibility does not depend on dephasing and, in this plot, it is represented by a horizonal line that coincides with $\omega^2 \Re[\Delta \hat{\chi}^{(1)}(\omega)]$ in the limit $\omega \to 0$.
Intraband dynamics dominate the optical response for photon energies $\hbar\omega \lesssim 3$~eV, with the exception of the range $0.5\,\text{eV} \lesssim \hbar\omega \lesssim 0.6\,\text{eV}$.
The resonant transitions in this range mainly take place among the lowest conduction bands, which are degenerate at the $\Gamma$ point.
\begin{figure}
	\includegraphics[width=0.9\columnwidth]{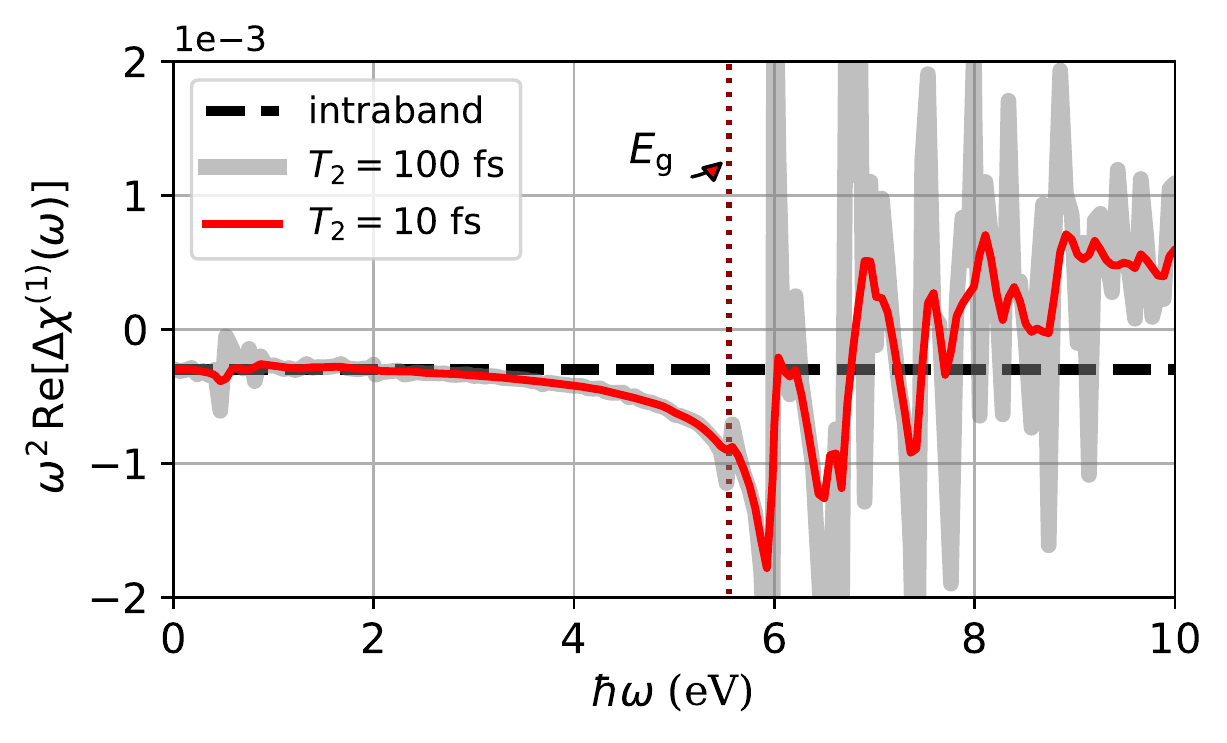}
	\caption{\label{fig:intra_vs_total_spectrum}
		The change of the real part of the linear susceptibility induced by a 1-V/{\AA} 4-fs 800-nm laser pulse in diamond (LDA).
		The dashed black line represents $\omega^2 \Delta \chi^{\mathrm{intra}}(\omega)$.
		Plotting $\omega^2 \Re[\Delta \chi^{(1)}(\omega)]$, we compare two dephasing times: $\gamma^{-1} = T_2 = 10$~fs (thin red curve) and $T_2 = 100$~fs (thick gray curve).
		The vertical dotted line shows the position of the band edge for direct transitions.
	}
\end{figure}

Figures \ref{fig:total_intra} and \ref{fig:inter_intra} show how the intra- and interband contributions to $\Re[\Delta \hat{\chi}^{(1)}(\omega)]$ depend on the peak laser field.
\begin{figure}
	\includegraphics[width=0.9\columnwidth]{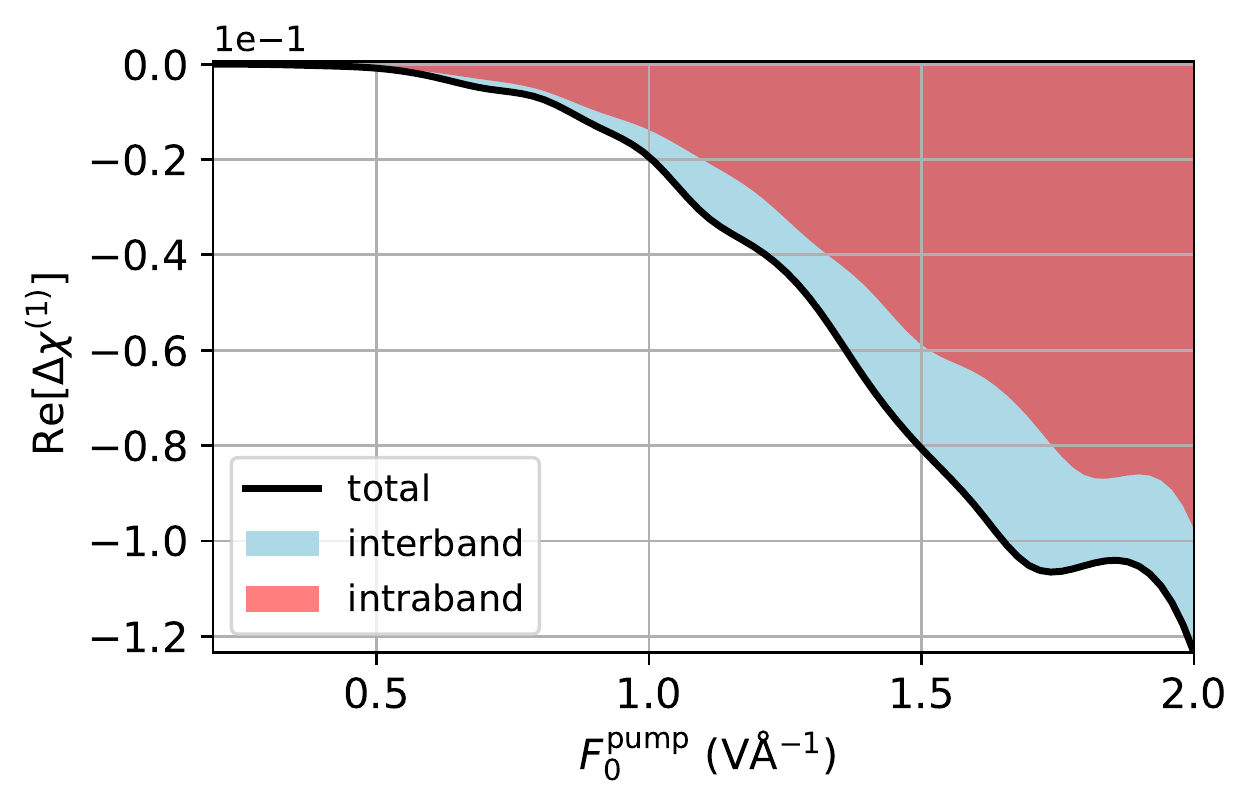}
	\caption{\label{fig:total_intra}
		The change of the linear response at $\hbar\omega = 4$~eV as a function of the peak laser field for $\gamma^{-1} = T_2 = 10$~fs.
	}
\end{figure}
Since we consider the case where all energy bands are initially either fully occupied or empty, both $\Delta \hat{\chi}^{\mathrm{intra}}$ and $\Delta \hat{\chi}^{\mathrm{inter}}$ are proportional to $N_{e-h}$ and, therefore, rapidly increase with the amplitude of the laser pulse as shown in Fig.~\ref{fig:total_intra}.
Furthermore, both $\Delta \hat{\chi}^{\mathrm{intra}}$ and $\Re[\Delta \hat{\chi}^{\mathrm{inter}}]$ are negative at $\hbar\omega = 4$~eV.
\begin{figure}
	\begin{tabular}{p{0.95\columnwidth} p{0pt}}
		\vspace{0mm} \includegraphics[width=0.95\columnwidth]{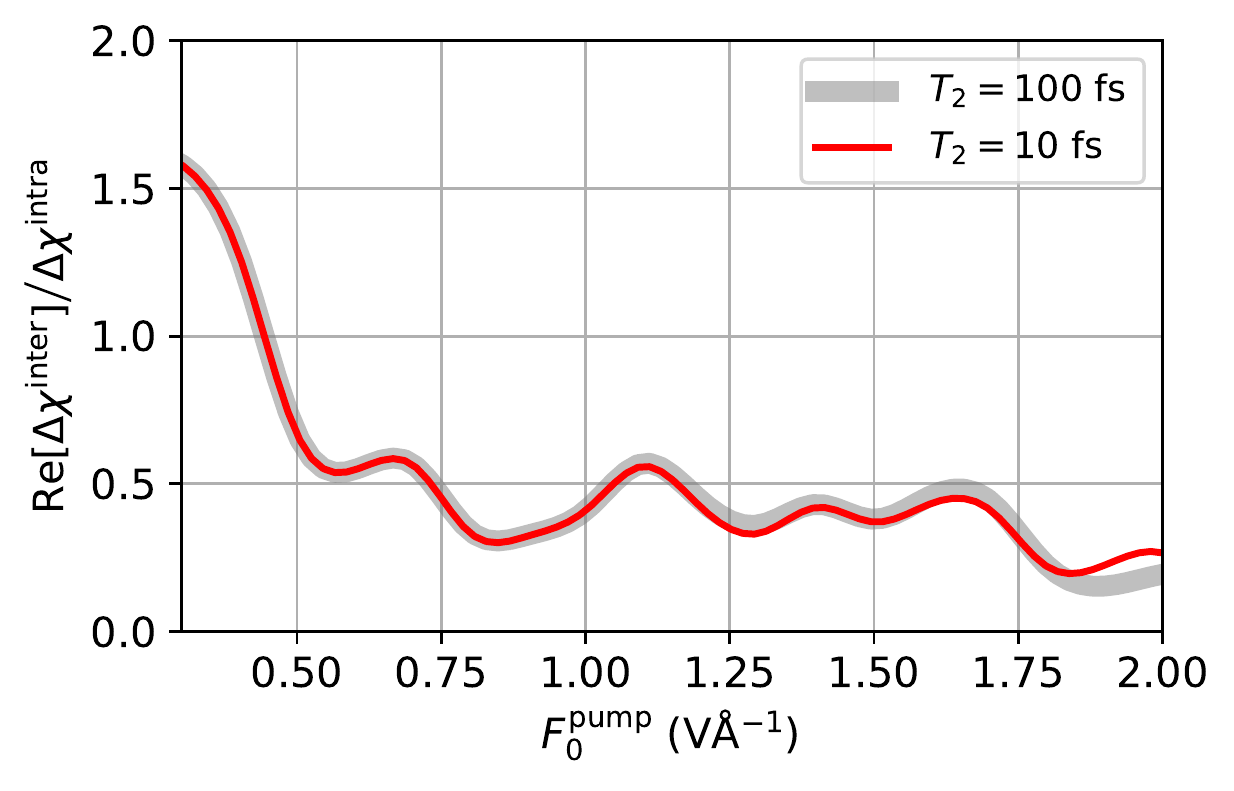} &
		\vspace{1mm} \hspace{-1.0\columnwidth}
		\textbf{(a)}
	\end{tabular}\\[-5mm]
	\begin{tabular}{p{0.95\columnwidth} p{0pt}}
		\vspace{0mm} \includegraphics[width=0.95\columnwidth]{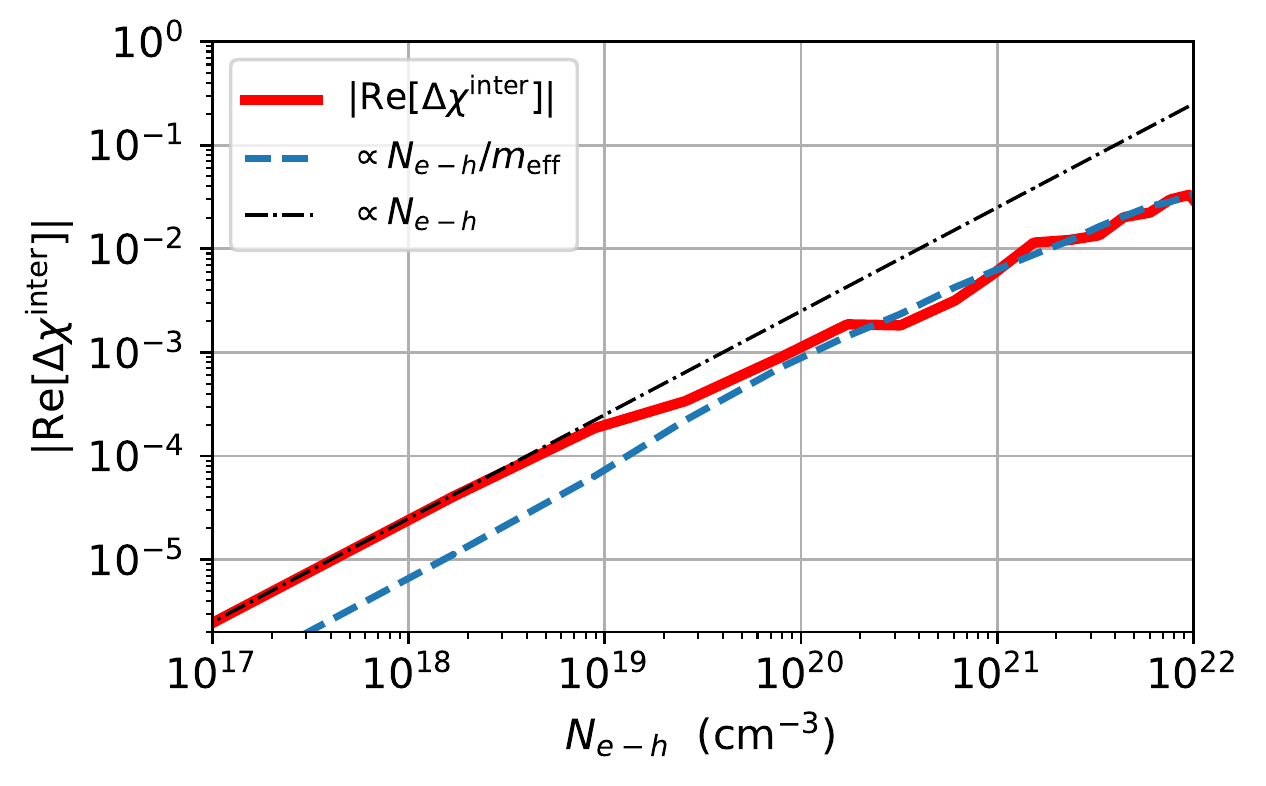} &
		\vspace{1mm} \hspace{-1.0\columnwidth}
		\textbf{(b)}
	\end{tabular}
	\caption{\label{fig:inter_intra}
		The laser-induced susceptibility change, $\Delta \chi^{(1)}(\omega) \equiv \mathbf{e}_{\mathrm{probe}} (\Delta \hat{\chi}^{(1)}(\omega)\,\mathbf{e}_{\mathrm{probe}})$, at $\hbar\omega = 4$~eV.
		(a) The ratio of $\Re[\Delta \chi^{\mathrm{inter}}]$ to the intraband component of $\Delta \chi^{(1)}$, plotted against the peak laser field.
		Consistently with Fig.~\ref{fig:intra_vs_total_spectrum}, there is almost no difference between the two dephasing times: $\gamma^{-1} = T_2 = 10$~fs (thin red curve) and $T_2 = 100$~fs (thick gray curve).
		(b) The dependence of $\Re[\Delta \chi^{\mathrm{inter}}]$ on the concentration of charge carriers for $T_2 = 10$~fs.
	}
\end{figure}

It is well known that band curvatures control the probabilities of interband transitions.
In particular, this is why the effective mass appears in the Keldysh parameter~\cite{Keldysh_JETP_1965}.
Figure \ref{fig:inter_intra} strongly suggests that, for large excitation probabilities, the relationship between the average band curvatures and the interband component of the laser-induced susceptibility change is particularly simple: $\Re[\Delta \chi^{\mathrm{inter}}] \propto N_{e-h}/m_{\mathrm{eff}}$.
This result is not obvious because the effective mass does not explicitly appear in Eq.~\eqref{eq:chi_inter}.
Also, for weak laser fields, $\Re[\Delta \chi^{\mathrm{inter}}] \propto N_{e-h}$.
The change of the scaling law as $F_0^{\mathrm{pump}}$ increases explains the decrease of $\Re[\Delta \chi^{\mathrm{inter}}] / \Delta \chi^{\mathrm{intra}}$ in Fig.~\ref{fig:inter_intra}(a).
(Note that $\Re[\Delta \hat{\chi}^{\mathrm{intra}}] \propto N_{e-h}/m_{\mathrm{eff}}$ by definition.)
The observation that $\Re[\Delta \chi^{\mathrm{inter}}] \propto \Delta \hat{\chi}^{\mathrm{intra}}$ in the strong-field regime is consistent with the interdependence of inter- and intraband dynamics~\cite{Wismer_PRL_2016, Schlaepfer_NaturePhysics_2018, Kruchinin_RMP_2018}.
It also explains why the Drude fit works so well~\cite{Sato_PRB_2014_silicon} for the net response, $\Re[\Delta \chi^{(1)}]$.
The fact that a substantial part of this response is of the interband nature translates into the phenomenological relaxation time, which was found to be on the order of $\sim 1$~fs in numerical simulations that neglected relaxation processes~\cite{Yabana_PRB_2012}.


\subsection{Excitation-induced birefringence}

Prior to the excitation by a laser pulse, diamond is an isotropic crystal.
By exciting charge carriers, a laser pulse induces birefringence, which is easily measured by optical means.
One could expect that a linearly polarized pulse should turn diamond into a uniaxial crystal, but, according to our calculations, this is not the case---for a sufficiently strong injection field, laser-excited diamond is, in general, a biaxial crystal.
In Fig.~\ref{fig:birefringence}, we illustrate the induced birefringence probed by an infinitesimally weak pulse.
We obtained the data for this figure by analyzing the $\hat{\chi}^{(1)}(\omega)$ tensor [see Eqs.~\eqref{eq:chi_intra} and \eqref{eq:chi_inter}]. 
For a given wave vector, $\mathbf{k}$, a biaxial crystal supports two modes characterized by effective permittivities, $\varepsilon_\mathrm{eff}$, which satisfy the following equation:
\begin{equation}
	\text{det} \left[ \mathbb{1} + 4 \pi \hat{\chi}^{(1)} - \varepsilon_\mathrm{eff}
	\left( \mathbb{1} - \frac{\mathbf{k} \mathbf{k}^{\top}}{\lVert \mathbf{k} \rVert^2} \right) \right] = 0.
\end{equation}
Birefringence consists in the two effective permittivities being different.

\begin{figure}
	\begin{tabular}{p{0.95\columnwidth} p{0pt}}
		\flushright
		\vspace{0mm} \includegraphics[width=0.95\columnwidth]{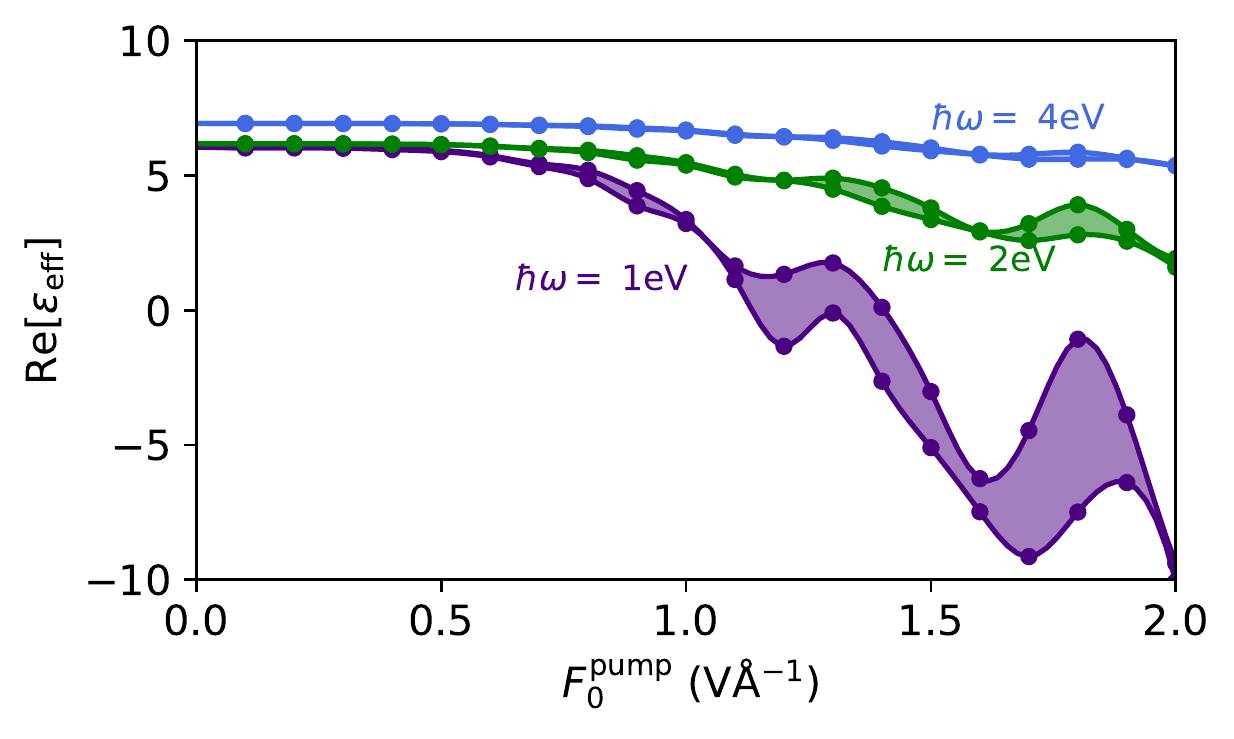} &
		\vspace{1mm} \hspace{-1.0\columnwidth}
		\textbf{(a)}
	\end{tabular}\\[-8mm]
    \begin{tabular}{p{0.95\columnwidth} p{0pt}}
      \flushright
	  \vspace{0mm} \includegraphics[width=0.925\columnwidth]{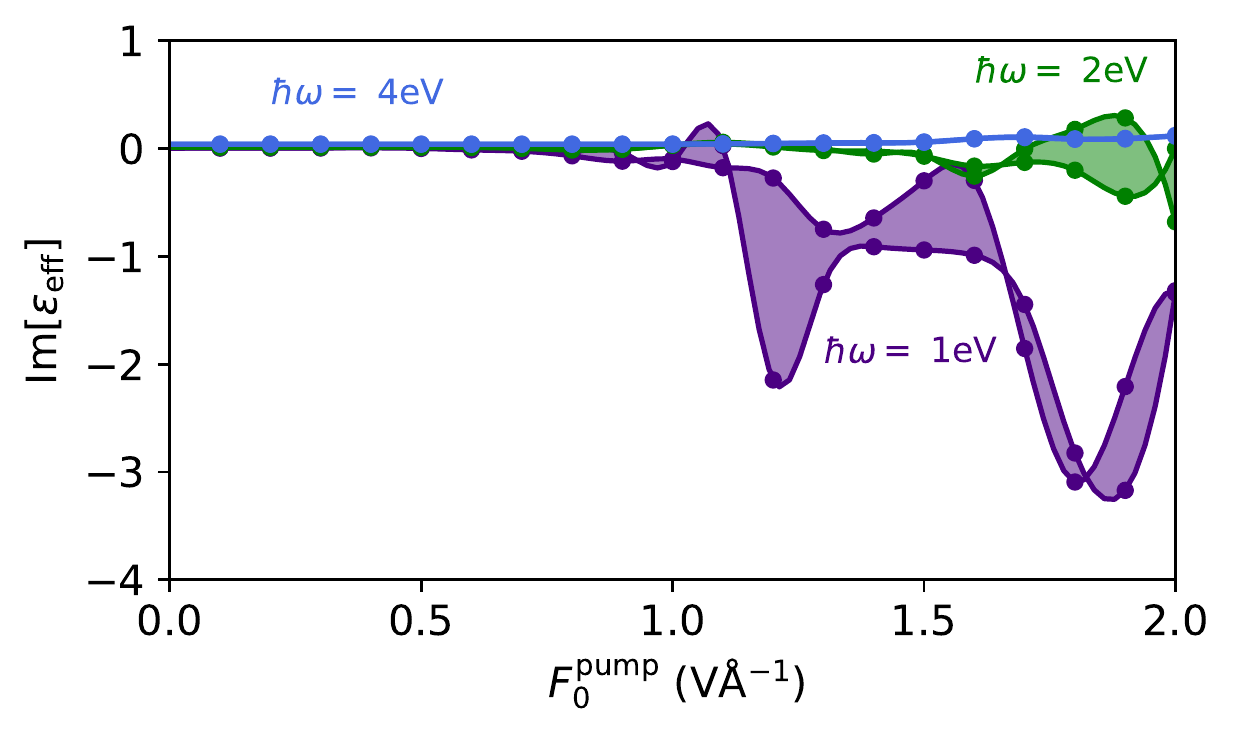} &
	  \vspace{1mm} \hspace{-1.0\columnwidth}
	  \textbf{(b)}
    \end{tabular}
	\caption{\label{fig:birefringence}
		The effective permittivities for light propagating in the direction of the pump beam.
		The three shaded areas correspond to light frequencies $\hbar\omega = 1$~eV (magenta), 2~eV (green), and 4~eV (blue).
		The upper and lower boundaries of each shaded area correspond to two modes that propagate preserving their polarization state (the polarization directions of these modes depend on $F_0^{\mathrm{pump}}$).
		The dots represent numerical data, the curves are cubic splines.
		For these calculations, we used $\gamma^{-1} = T_2 = 10$~fs.
	}
\end{figure}
In Fig.~\ref{fig:birefringence}, we plot the effective permittivities for three frequencies of probe light: $\hbar\omega \in \{1, 2, 4\}$~eV, which we show in magenta, green, and blue, respectively.
We see from this plot that the induced birefringence decreases with the probe frequency and increases with the strength of the injection field.
The dependence on $F_0^{\mathrm{pump}}$ is, however, not monotonous.
We observe particularly large values of the induced birefringence for field strengths where the real part of the permittivity takes negative values due to the presence of electron--hole plasma.
This happens for plasma frequencies  $\omega_{\mathrm{pl}} = \sqrt{4 \pi N_{e-h} e^2 / m_{\mathrm{eff}}} \gtrsim \omega n_0(\omega)$, where $n_0$ is the unperturbed refractive index of the solid.
For the $\hbar\omega = 1$~eV dataset, this condition is fulfilled for $F_0^{\mathrm{pump}} \gtrsim 1.2$~V/{\AA}.

\subsection{Pump--probe simulations}
\label{sec:pump-probe}
This subsection provides evidence that the average effective mass defined by Eq.~\eqref{eq:average_inverse_mass} indeed determines the strength of the electric current induced by a probe pulse. 
\begin{figure}
	\includegraphics[width=0.9\columnwidth]{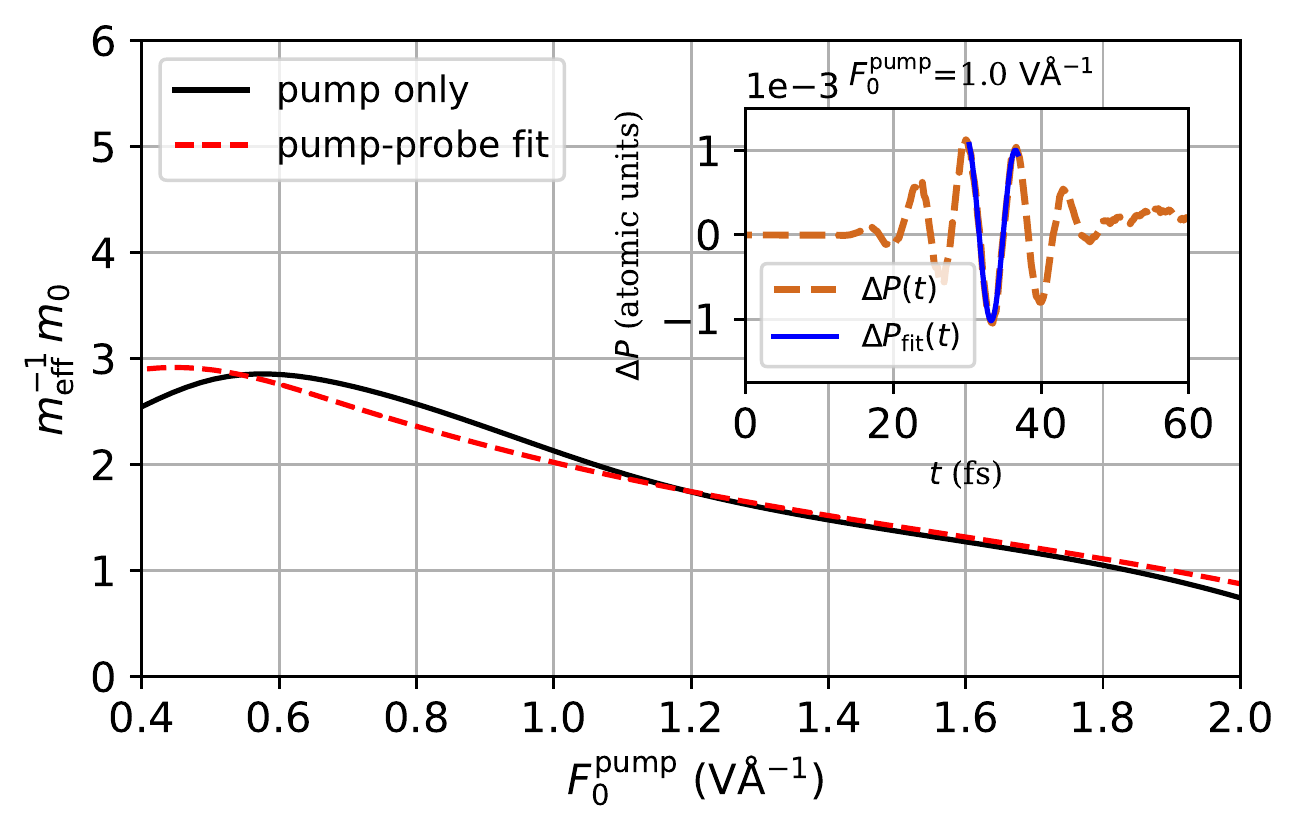}
	\caption{\label{fig:fit_mass}
		Comparison of the effective inverse masses calculated with two methods:
		The solid black curve, labeled as ``pump only'', is identical to that in Fig.~\ref{fig:electrons_vs_holes}; it was obtained by applying Eq.~\eqref{eq:average_inverse_mass} to the outcomes of simulations with a sole pump pulse.
		The dashed red curve represents pump--probe simulations, where the time-dependent polarization induced by the probe pulse was approximated with the relaxation-free Drude model.
		The inset illustrates the Drude fit for $F_0^\mathrm{pump} = 1$~V/{\AA} and $F_0^\mathrm{probe} = 0.1$~V/{\AA}.
	}
\end{figure}
For these pump--probe simulations, we used the same 4-fs 800-nm pump pulse, followed by a 12-fs 2000-nm linearly polarized probe pulse with a peak electric field of 0.1~V/{\AA}.
The two pulses had no overlap, and their polarizations were orthogonal to each other.
To determine the effective mass from the pump--probe simulations, we first evaluate how the pump pulse changes the medium polarization in the direction of the probe field.
We accomplish this by subtracting the polarization induced by the sole probe pulse from that induced by both pulses.
For each amplitude of the pump pulse, this procedure yields a time-dependent function $\Delta P(t)$, which we fit with the following ansatz within the central cycle of the probe pulse:
\begin{equation}
	\Delta P(t) \approx P_0 + J_0 t + e^2 N_{e-h} m_{\mathrm{eff}}^{-1} \int_t^{\infty} A_{\mathrm{probe}}(t')\,\de t'.
\end{equation}
Here, $N_{e-h}$ is the concentration of conduction-band electrons after the pump pulse.
From the three fit parameters ($P_0$, $J_0$, and $m_{\mathrm{eff}}^{-1}$), we are interested only in $m_{\mathrm{eff}}^{-1}$, plotting it with the dashed curve in Fig.~\ref{fig:fit_mass}.
The good agreement between the outcomes of this analysis and those of Eq.~\eqref{eq:average_inverse_mass} validates the analysis presented in the previous subsections.
This is not a trivial result because the assumption of purely intraband motion is generally inapplicable at those crystal momenta where some bands are either degenerate or experience an avoided crossing.
In the vicinity of such crystal momenta, even a weak infrared pulse can drive interband transitions with a significant probability~\cite{Yakovlev_Springer_2016}.
Figure~\ref{fig:fit_mass} demonstrates that even though the dynamics of a particular charge carrier in the field of a weak probe pulse may violate our assumptions, the dynamics of the entire electron--hole plasma are well described by the intraband approximation, especially when charge carriers occupy a large part of the Brillouin zone.
The good agreement illustrated by Fig.~\ref{fig:fit_mass} also demonstrates that interband coherences, neglected in Eqs.~\eqref{eq:chi_intra} and \eqref{eq:chi_inter}, have a negligible effect on the low-frequency optical response of a laser-excited solid.

The inset in Fig.~\ref{fig:fit_mass} illustrates how the fit was performed.
We also note that $\Delta P$ after the probe pulse is not zero.
Charge carriers can get displaced and accelerated by the end of a weak probe pulse, which would be impossible if the pulse induced strictly intraband dynamics.
The transitions that are responsible for the formation of the residual polarization and electric current are the same transitions that manifest themselves in Fig.~\ref{fig:intra_vs_total_spectrum} as the low-energy resonances.

The effective mass evaluated from the Drude fit depends on the amplitude of the probe pulse, which we illustrate in Fig.~\ref{fig:fit_mass_probe}.
We obtained this data by scanning over $F_0^{\mathrm{probe}}$ in the same pump--probe arrangement as before.
The amplitude of the pump pulse was $F_0^\mathrm{pump} = 1$~V/{\AA}.
The decrease of $m_{\mathrm{eff}}^{-1}$ with a growing amplitude of the probe pulse is due to the interband motion in nonparabolic bands.
Indeed, for $F_0^{\mathrm{probe}} = 0.4$~V/{\AA}, the amplitude of reciprocal-space excursion is $|e F_0^{\mathrm{probe}}| / (\hbar \omega_{\mathrm{probe}}) = 6.45\ \text{nm}^{-1}$, which is as large as 37\% of the reciprocal-lattice period.
The fact that the effective mass considerably depends on the amplitude of the probe pulse implies a significant nonlinearity of the intraband polarization response.
\begin{figure}
	\includegraphics[width=0.9\columnwidth]{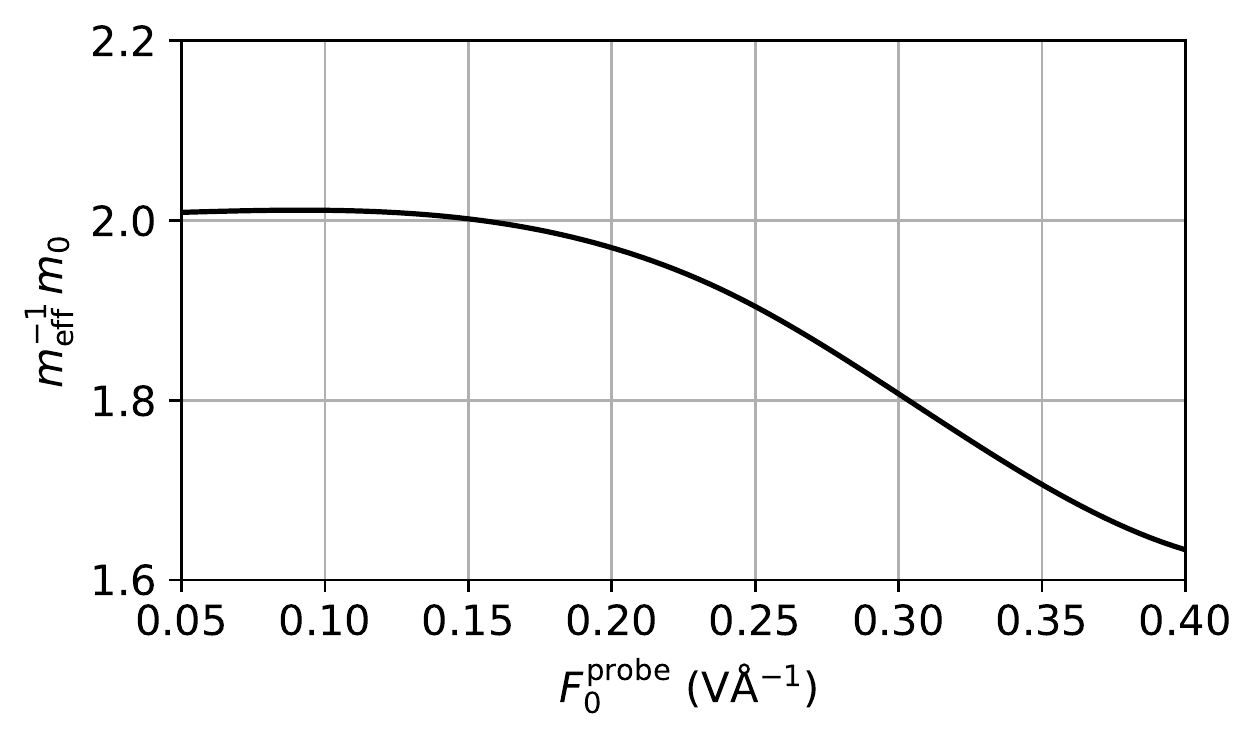}
	\caption{\label{fig:fit_mass_probe}
		The dependence of the effective inverse mass, evaluated from the Drude fit, on the amplitude of the probe pulse.
		For these simulations, we used $F_0^\mathrm{pump} = 1$~V/{\AA}.
	}
\end{figure}

\section{Summary}
The effective mass averaged over all charge carries excited by a laser pulse in a transparent solid strongly depends on the amplitude of the pulse.
This effect stems from band nonparabolicity, it is particularly important in the tunneling regime, and it is pronounced in all the solids that we investigated.
Apart from pointing out the magnitude of this effect, we also make several observations related to its nature and properties. 
Even though the coherence between energy bands occupied by a pump pulse has measurable outcomes~\cite{Mashiko_Nature-Physics_2016}, we point out that it has a minor effect on the permittivity change within the transparency region.
This is one of the reasons why the Drude response dominates $\Delta\hat{\chi}^{(1)}(\omega)$ in most of this region.
The insignificance of interband coherences means that, to a good approximation, an average effective mass depends only on band occupations and band curvatures.
Moreover, the availability of $\mathbf{k}$-dependent data is not essential for evaluating the average effective mass---it can be estimated with a reasonable accuracy from energy-dependent average band curvatures, excitation probabilities, and the density of states.
It is possible because the average mass of charge carriers within an eV-broad energy range weakly depends on the peak electric field of a laser pulse.

We observed that, starting from a certain field strength, the interband component of $\Delta\hat{\chi}^{(1)}(\omega)$ becomes proportional to the $N_{e-h}/m_{\mathrm{eff}}$ ratio, that is, to the square of the plasma frequency; in this regime, we expect the intraband motion to have an impact on interband transitions.
Investigating the excitation-induced birefringence, we observed that it is particularly large when the plasma frequency exceeds the probe-pulse frequency multiplied by the unperturbed refractive index.
We also observed that the average effective mass that describes the ballistic acceleration of charge carriers in the field of a near-infrared probe pulse considerably depends on the pulse's amplitude.

\begin{acknowledgments}
Supported by the DFG Cluster of Excellence: Munich-Centre for Advanced Photonics.
M.\,S.\,W.\ was supported by the International Max Planck Research School of Advanced Photon Science (IMPRS-APS).
The authors thank Claudio Attaccalite for his advice on using DFT codes.
\end{acknowledgments}


%

\end{document}